# Model predictive control for the prescription of antithyroid agents

**M. Menzel[1]\*, T. M. Wolff[1], J. W. Dietrich[2,3,4], and M. A. Müller[1]**

[1]Leibniz University Hannover, Institute of Automatic Control, Germany
[2]Diabetes, Endocrinology and Metabolism Section, Department of Internal Medicine I, St. Josef Hospital, Ruhr University Bochum, Germany
[3]Diabetes Centre Bochum-Hattingen, St. Elisabeth-Hospital Blankenstein, Hattingen, Germany
[4]Ruhr Center for RareDiseases (CeSER), Ruhr University of Bochum and Witten/Herdecke University, Bochum, Germany
\* Corresponding author, email: maylin.menzel@stud.uni-hannover.de

*Abstract: Although hyperthyroidism is a common disease, the pharmaceutical therapy is based on a trial-and-error approach. We extend a mathematical model of the pituitary-thyroid feedback loop such that the intake of one antithyroid agent, namely methimazole (MMI), can be considered and use a model predictive control (MPC) scheme to determine suitable dosages.*

## I. Introduction

Thyroid disorders are widespread medical conditions. The cause for these conditions is often related to a disturbance of the pituitary-thyroid feedback loop, an important control loop of the endocrine system. Under normal circumstances, the pituitary gland releases thyroid-stimulating hormone ($TSH$) upon stimulation of thyrotropin-releasing hormone ($TRH$). Inside the thyroid gland, a stimulation of $TSH$ leads to the production and secretion of the thyroid hormones triiodothyronine ($T_3$) and thyroxine ($T_4$). In turn, $T_4$ inhibits the release of $TSH$ from the pituitary gland.

In case of hyperthyroidism, the thyroid gland produces too much $T_3$ and/or $T_4$ and the increased hormone concentrations lead to symptoms like heart palpitation, weight loss, and goiter and may be even life-threatening in the case of thyroid storm. Treatment of hyperthyroidism aims to reduce the thyroid function and normalize the hormone concentrations. The current treatment options for patients are thyroidectomy, radioiodine therapy and antithyroid agents. Definitive therapy reduces the amount of functional thyroid tissue by either removing parts of it or damaging it with radioactive radiation. These forms of therapy are irreversible and apart from the obvious risks of surgery and exposure to radioactive radiation, they often lead to hypothyroidism. Antithyroid agents, which are necessary for pretreatment of definitive therapy, reversibly inhibit the production of thyroid hormones and are the only treatment option that rarely induces permanent hypothyroidism.

## II. Model extensions and controller design

The results discussed within this abstract are based on [1]. In order to render this work as self-contained as possible, in the following we recall the details of the model extensions and the controller design from [1].

The mechanisms of the pituitary-thyroid feedback loop are modeled in [1] as a system of six nonlinear first order differential equations which describe different hormone concentrations. To model hyperthyroidism, we increase the secretory capacity of the thyroid gland (named $G_T$) in the first term of the differential equation of $T_{4,th}$, i.e., (compare also [1, Eq. (A.1)])

$$G_T \frac{TSH(t)}{TSH(t) + D_T}. \quad (1)$$

The antithyroid agent $MMI$ decreases the production of thyroid hormones by inhibiting the enzyme thyroid peroxidase ($TPO$) which catalyzes an important step in the production process, the conversion of inorganic iodide ($I_C$) to organic bound iodide ($I_{Tg}$). To consider this mechanism, we multiply the term in equation (1) of the differential equation related to $T_{4,th}$ with the state $I_{Tg}$. The differential equation for $I_{Tg}$ is shown in [1, Eq. (A.7)].

To model the effect of $MMI$ inside the thyroid gland, we first have to determine the plasma concentration $MMI_{Plas}$ after an $MMI$ intake. In this abstract, we consider the most common form of application, which is the oral intake of $MMI$. The plasma concentration resulting from a single orally taken dosage $u(t_0)$ can be described as

$$MMI_{Plas}(t) = \frac{u(t_0)k_a f}{V(k_a - k_e)}(e^{-k_e t} - e^{-k_a t}). \quad (2)$$

According to [2], the bioavailability $f$ of $MMI$ is 93 %. The remaining parameters denote the volume of distribution $V$ as well as the elimination constant $k_e$ and the absorption constant $k_a$, $V = 28.8$ L, $k_e = 0{,}1857$ h$^{-1}$ and $k_a = 11$ h$^{-1}$ based on [3].

Next, we determine the resulting concentration of $MMI_{th}$, denoting the intrathyroidal $MMI$ concentration. We choose heuristically

$$\frac{dMMI_t}{dt}(t) = MMI_2(t) \quad (3)$$

$$\frac{dMMI_2}{dt}(t) = -a_0 MMI_1(t) - a_1 MMI_2(t) + MMI_{Plas}(t) \quad (4)$$

$$MMI_{th}(t) = b_0 MMI_1(t) + b_1 MMI_2(t) \quad (5)$$

to model this process. The parameters are estimated in a least-squares sense based on data from [4] and result in $b_1 = 690.3 \cdot 10^{-6}$, $b_0 = 37 \cdot 10^{-9}$, $a_1 = 92.2 \cdot 10^{-6}$ and $a_0 = 2.5 \cdot 10^{-9}$.

Third, we determine the remaining activity of $TPO$ ($TPO_a$) in relation to the concentration of $MMI_{th}$ as well as its substrate $I_C$. To this end, we choose heuristically

$$TPO_a(t) = c_0 \left(1 + exp\left(-c_1\left(-MMI_{th}(t)^{-\frac{1}{c_2}} + c_3\right)\right)\right)^{-1}. \quad (6)$$

and identify the parameters in a least-squares sense based on data from [5]. When considering a typical concentration of $I_C$, the resulting values are $c_0 = 0.9$, $c_1 = 84.1 \cdot 10^3$, $c_2 = 1.3$ and $c_3 = 80.5 \cdot 10^{-6}$. An increased intrathyroidal $I_C$ concentration which could, e.g., occur after the intake of contrast agents for radiographic imaging, results in the parameters $c_0 = 1$, $c_1 = 175.8 \cdot 10^3$, $c_2 = 5$ and $c_3 = 97.6 \cdot 10^{-3}$. Then, we multiply equation (6) with $TPO(t)$ within the state equation of $I_{Tg}$.

Finally, we implement a model predictive control (MPC) scheme to determine the dosages of $MMI$. At each sampling instance $t$ (with $t \in \mathbb{N}_0$), we measure the system's state $x(t)$ and determine the optimal input sequence over some control horizon $T$, in this abstract 15 days, with respect to a cost function. We then apply the first element in the optimal sequence to the system. The cost function chosen for this abstract penalizes deviations from the targeted state, variations from the last input and the magnitude of the input. Additionally, we consider input constraints to limit the maximal dosage. A more detailed mathematical description of the MPC can be found in [1].

## III. Simulation Results

We simulate two different clinically relevant scenarios. For each of these, we execute two simulations, once for a nominal case (dotted lines) and a second time for a case with disturbances (continuous lines). In the latter, we include a measurement noise and an exemplary model-plant mismatch. The added measurement noise follows a Gaussian distribution with $\mu = 0$ and $\sigma = 0.11$ and is truncated at $\pm 0.3$ to avoid negative hormone concentrations. To simulate the model-plant mismatch, we increase the values of the maximum activity of 5'-deiodinase type I ($G_{D1}$) and the maximum activity of the direct $T_3$ synthesis ($G_{T3}$) by 15% and the value of the maximum activity of 5'-deiodinase type II ($G_{D2}$) by 5% (see [1, Eq. (1)], [1, Eq. (A.3)], and ([1, Eq. (A.4)]). Furthermore, we consider disturbances in the form of forgotten dosages at days 4, 12 and 34 as well as accidental doubled intakes at days 13 and 26. The targeted hormone concentrations are represented by dashed lines. In comparison to [1], we simulate a more challenging and practically relevant situation. Here, we consider a higher noise level, a more severe model-plant mismatch, accidental doubled dosages, as well as forgotten dosages in the steady state.

Fig. 1 represents a normal case of hyperthyroidism where the patient is treated with one daily oral dosage. In Fig. 2,

we consider a patient with constant increased levels of iodide concentrations, where the patient also takes in one oral dosage per day. The systems start in their hyperthyroid steady states.

In both cases, the determined dosages are in line with clinical guidelines, e.g., that higher $MMI$ dosages are necessary to normalize the hormone concentrations in the case of increased iodide concentrations [6].

## IV. Conclusion

In this abstract, we determined optimal medication dosages for the treatment of hyperthyroidism with a model of the pituitary-thyroid feedback loop and an MPC yielding promising results. Currently, we assume that all states are measurable, which is not the case in clinical practice. Therefore, an interesting subject of future research is the implementation of a nonlinear observer.

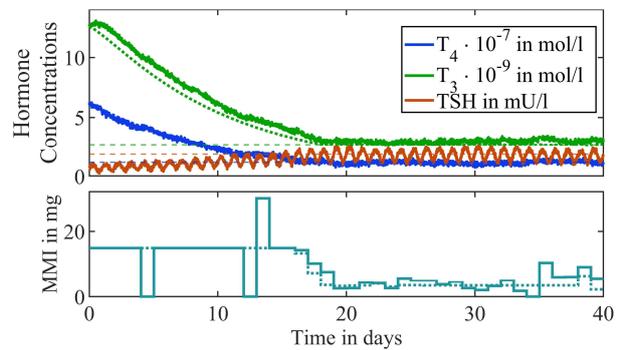

*Figure 1: Simulation results for one daily oral MMI intake.*

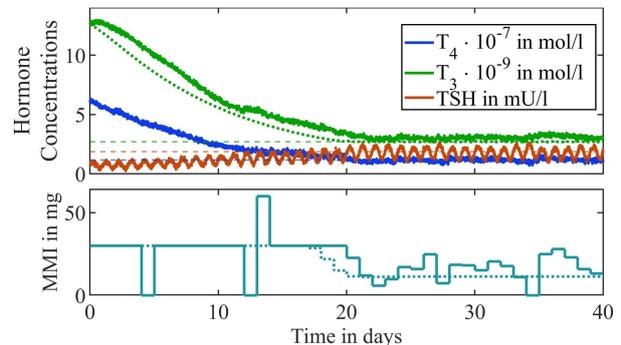

*Figure 2: Simulation results for one daily MMI intake and increased intrathyroidal iodide concentration.*

**AUTHOR'S STATEMENT**
This project has received funding from the European Research Council (ERC) under the European Union's Horizon 2020 research and innovation programme (grant agreement No 948679).